\documentclass[twocolumn,showpacs,amsmath,amssymb]{revtex4-1}
\usepackage{graphicx}
\usepackage{epstopdf}
\usepackage{subfigure}

\usepackage{dcolumn}
\usepackage{bm}

\DeclareMathOperator{\Tr}{Tr}
\newcommand{\kb}[0]{k_\text{B}}

\begin{document}

\title{Determination of the Origin and Magnitude of Logarithmic Finite-Size Effects on Interfacial Tension: Role of Interfacial Fluctuations and Domain Breathing}

\author{Fabian Schmitz}
\author{Peter Virnau}
\author{Kurt Binder}
\affiliation{%
\textit{ Institut f\"{u}r Physik, Johannes Gutenberg-Universit\"{a}t} \\
                   \textit{D-55099 Mainz, Staudinger Weg 7, Germany}
}%

\begin{abstract}
The ensemble-switch method for computing wall excess free energies of condensed matter is extended to estimate the interface free energies between coexisting phases very accurately. By this method, system geometries with linear dimensions $L$ parallel and $L_z$ perpendicular to the interface with various boundary conditions in the canonical or grandcanonical ensemble can be studied. Using two- and three-dimensional Ising models, the nature of the occurring logarithmic finite size corrections is studied. It is found crucial to include interfacial fluctuations due to ``domain breathing''.
\end{abstract}

\pacs{64.70.F-, 68.03.Cd, 64.60.an, 64.60.De}

\maketitle

Interfaces between coexisting phases occur in many contexts, nucleation of ice or water droplets in the atmosphere \cite{1,2}, hadron condensation from the quark-gluon plasma \cite{3}, etc. Interfacial free energies are driving forces for phase separation kinetics (droplet coarsening) \cite{4}, microfluidic processes \cite{5}, wetting and spreading \cite{6,7,8}, and capillary condensation or evaporation \cite{9,10,11}. These phenomena are fascinating problems of statistical mechanics and have important applications (in nanoscopic devices, materials science of thin films and surfactant layers (e.g. \cite{12}) extracting oil and gas from porous rocks \cite{9}, etc.).

Thus, the theoretical prediction of interfacial free energies has been a longstanding problem (see \cite{13,14,15} for reviews). Mean-field type theories \cite{16,17,18} neglect interfacial fluctuations (capillary waves \cite{19,20,21}) and hence are unreliable. Exact solutions exist in exceptional cases only, e.g. the Ising model in $d=2$ dimensions \cite{22}. Most efforts to compute interfacial free energies use computer simulation (e.g. \cite{15,23,24,25,26,27,28,29,30,31,32,33,34,35}). However, often different variants of these methods yield estimates disagreeing with each other far beyond statistical errors, e.g., for the hard sphere liquid-solid interface tension discrepancies of about 10\% occur \cite{33,34,35,36}.

\begin{figure}[htbp]
\centering
\subfigure{\includegraphics[clip=true, trim=3mm 3mm 3mm 3mm, angle=0,width= 0.4 \columnwidth]{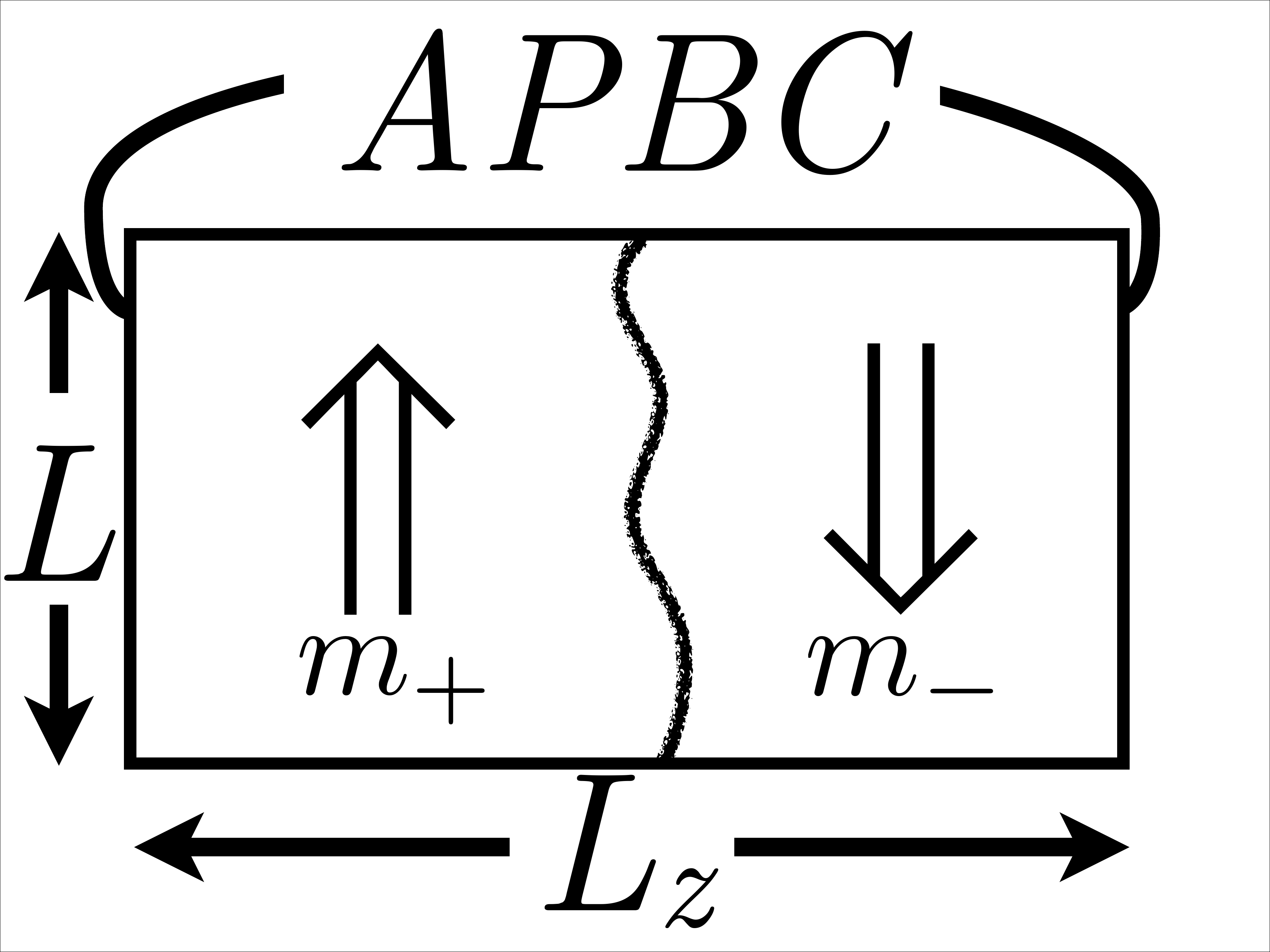}}
\subfigure{\includegraphics[clip=true, trim=3mm 3mm 3mm 3mm, angle=0,width= 0.4 \columnwidth]{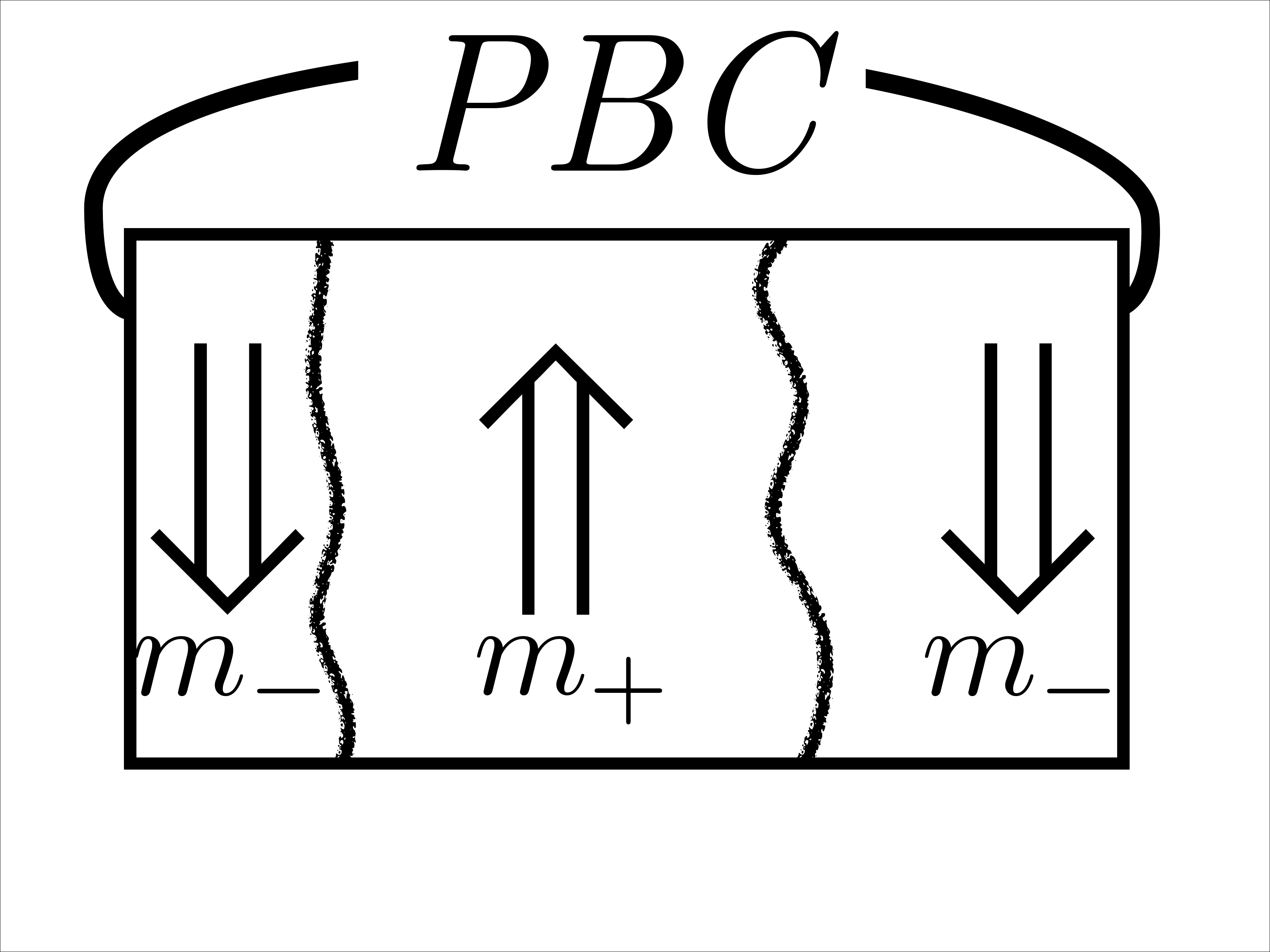}}
\caption{\label{fig1} Useful boundary conditions to study interfaces. For simplicity, we specialize to a $d$-dimensional Ising system in a box of linear dimension(s) $L$ parallel to the interface(s) [shown by thick wavy lines], and a linear dimension $L_z$ perpendicular to the interface. In parallel directions, periodic boundary conditions (PBC) are applied throughout. The double arrows indicate the sign of the magnetization $(m_+ > 0$, $m_- < 0)$ in the domains. $\langle m_+ \rangle =m_\text{coex}$, $\langle m_- \rangle= - m_\text{coex}$, $m_\text{coex}$ being the spontaneous magnetization of the Ising model. Left side shows antiperiodic boundary condition (APBC) in $z$-direction; right side shows PBC in $z$-direction (then necessarily two interfaces must occur).}
\end{figure}

Finite size effects are a possible source of systematic errors, but often are disregarded due to a lack of a generally accepted theoretical framework. But finite size effects on interfacial tensions are expected \cite{37,38,39,40,41,42,43} and also of physical interest for capillary condensation, nanoparticles, etc. These effects are subtle due to the anisotropy introduced by a (planar) interface: the linear dimension~$L$ parallel to the interface constrains the capillary wave spectrum; the linear dimension~$L_z$ in perpendicular $(z)$~direction affects interface translation as a whole. Also the choices of boundary conditions (Fig.~\ref{fig1}) and of statistical ensemble [e.g. canonical~(c) vs. grand-canonical~(gc)] matter.

\begin{figure}[htbp]
\centering
\subfigure{\includegraphics[clip=true, trim=2mm 2mm 2mm 2mm, angle=0,width= 0.80 \columnwidth]{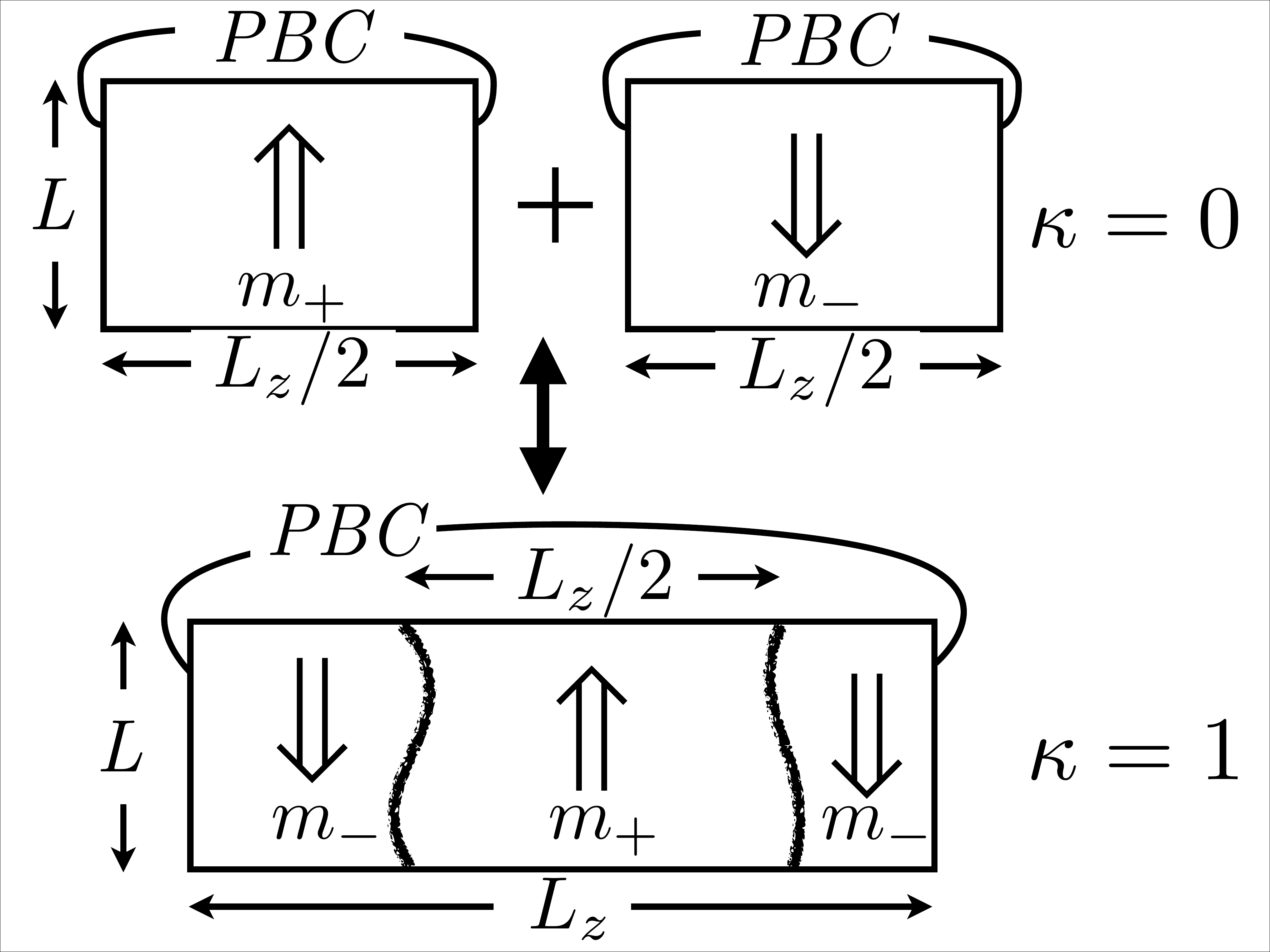}}
\subfigure{\includegraphics[clip=true, trim=25mm 0mm 20mm 0mm, angle=-90,width= 0.89 \columnwidth]{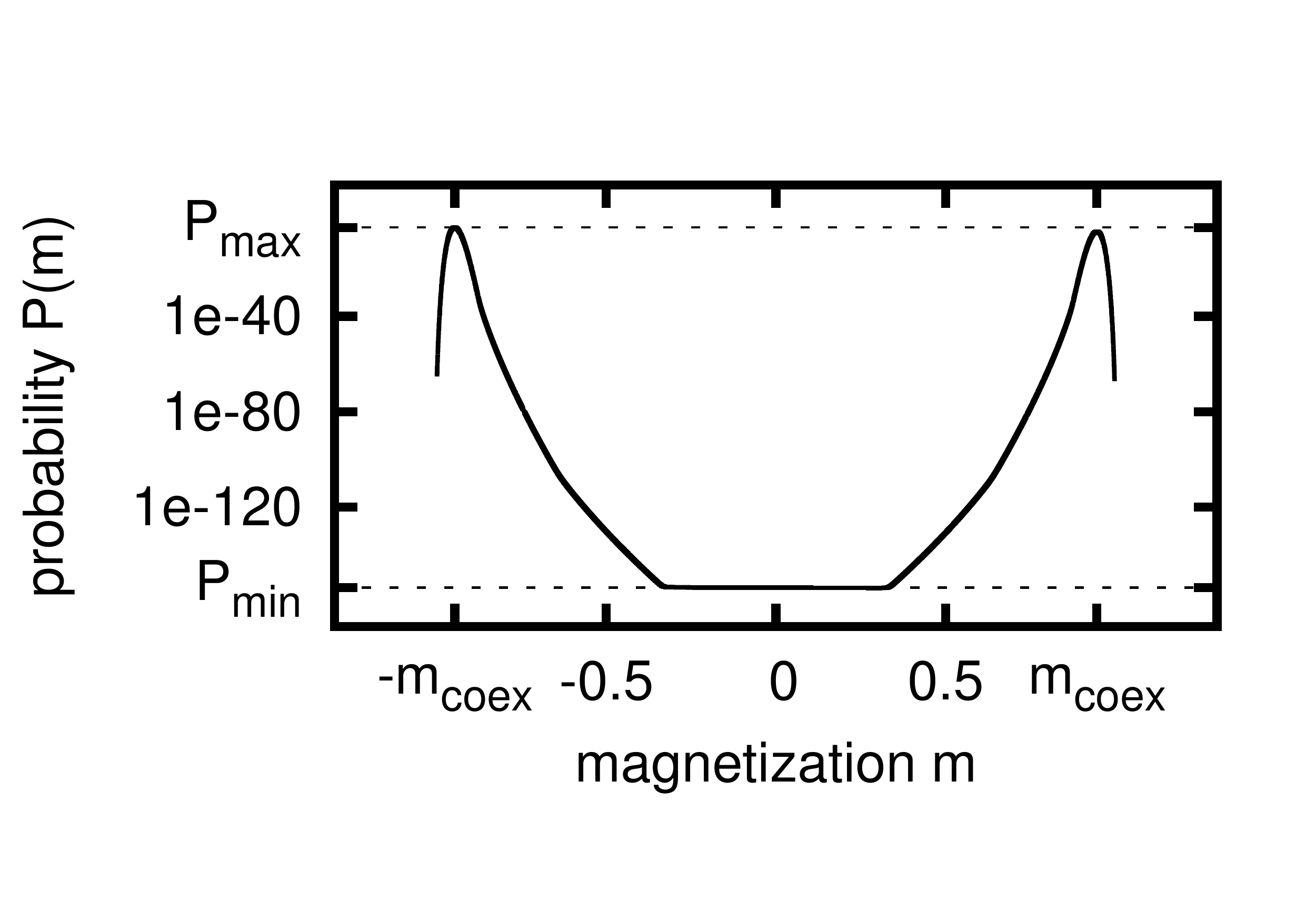}}
\caption{\label{fig2} (a) Schematic explanation of the ``ensemble switch method'' to find the interfacial free energy.
A system is constructed as a linear combination of two Hamiltonians $\mathcal{H} (\kappa) =\kappa \mathcal{H}_1 + (1 - \kappa) \mathcal{H}_0$, where $\mathcal{H}_1$ is the desired system of interest (containing interfaces). $\mathcal{H}_0$ consists of two separate systems of half the linear dimension $L_z$ each, and with PBC each so that no interfaces occur, and $0 \leq \kappa \leq 1$. The free energy difference between the states with $\mathcal{H}(\kappa=0)$ and $\mathcal{H}(\kappa=1)$ yields twice the interfacial free energy $\gamma$. (b) Sampling the probability distribution $P_{L,L_z}(m)$ where $m$ is the total magnetization per spin of a system with PBC throughout, one finds two sharp peaks (of height $P_\text{max}$) and a flat minimum (of height $P_\text{min}$) in between, with $\gamma = \ln P_\text{min}/(2 \kb T L ^{(d-1)})$. Data are for the case $d=3$, $L=20$, $\kb T/J=3.0$ Ising model.}
\end{figure}

This letter presents a discussion of these finite size effects affecting simulations and gives numerical evidence for the $d=2$ and $d=3$ Ising model for our theoretical results (that are believed to be of completely general validity). Our simulation evidence was made possible by extending the ``ensemble switch method'' \cite{44,45,46,47} for wall excess free energies to the computation of interfacial free energies (Fig.~\ref{fig2}). This new method is described next; at the outset we stress that this method is not restricted to systems possessing Ising-type symmetries between the coexisting phases.

The basic idea is to compute the free energy difference between two systems $(1,2)$ differing only by the absence $(1)$ or presence $(2)$ of interfaces (Fig.~\ref{fig2}a). Both systems have the same degrees of freedom and the same volume. System~1 is split along the $z$-direction into two halves (of linear dimensions $L_z/2)$, periodic boundary conditions are applied to each part individually. The left part is in the ``spin up'' phase ($m_+ > 0$), the right part in the ``spin down'' phase ($m_-< 0$), imposing the constraint for the total magnetization per spin $m=(m_+ + m_-)/2=0$. The same constraint applies to system~2, which contains two interfaces (consistent with the probability distribution $P_{L, L_z} (m)$, cf. Fig.~\ref{fig2}(b)). Thus, the Hamiltonian $\mathcal{H}_0$, $\mathcal{H}_1$ of the two systems $1,2$ differ only by the choice of boundary conditions. Using a parameter $\kappa$ with $0\leq \kappa \leq 1$, we define a Hamiltonian $\mathcal{H}({\kappa})=\kappa \mathcal{H}_1 + (1- \kappa) \mathcal{H}_0$, and the free energy $F(\kappa)=-\kb T \ln(\Tr \{\exp [-\mathcal{H}(\kappa)/\kb T\})$, $T$= absolute temperature, $\kb=$ Boltzmann's constant. The (dimensionless) interfacial tension then is
	\begin{equation} \label{eq1}
	\gamma_{L,L_z}=(2 L^{d-1} \kb T)^{-1} [F(1)-F(0)] \quad,
	\end{equation}
where $2L^{d-1}$ is the total interfacial area. This free energy difference is computed by thermodynamic integration, i.e.~dividing the interval for $\kappa$ into a number of discrete values $\kappa_i$ and considering Monte Carlo moves $\kappa_i \rightarrow \kappa_{i \pm 1}$, $F(\kappa_{i+1}) - F (\kappa_{i})$ is obtained via a parallelized version of successive umbrella sampling \cite{44,45,48}. On each core, the system can switch between two adjacent values $\kappa_i$ and $\kappa_{i+1}$. The logarithm of the ratio of the number of occurrences in two adjacent states corresponds to the difference in free energy. We expect -- and have verified -- that this method yields results equivalent to the estimates \cite{2,26,27} $\gamma_{L,L_z} =(2 \kb T L^{d-1})^{-1} \ln (P_\text{max}/P_\text{min})$ drawn from sampling $P_{L, L_z}$, cf. Fig.~\ref{fig2}b.

This new method has numerous advantages: (i) it can be applied to cases such as liquid-solid interfaces, for which probability distribution methods are difficult to apply \cite{32}. (ii) The generalization to antiperiodic (APBC, Fig.~\ref{fig1}a) or surface field (or fixed spin) boundary conditions is easy. In these cases, both the canonical ensemble ($m$ fixed, e.g. $m=0$) and the grand-canonical ensemble ($m$ freely fluctuates from about $-m_\text{coex}$ to about $+m_\text{coex}$) can be used. Note that for $\kappa=0$, the boundary conditions are always periodic in all directions, while for $\kappa=1$, one can use either PBC or APBC in $z$-direction. We will show that a comparative study of such choices is illuminating.

For $L$ and $L_z$ large enough, the leading finite size effects are described by two logarithmic corrections of opposite sign
	\begin{multline} \label{eq2}
	\gamma_{L,L_z} = \gamma_{\infty} - x_{\perp} \frac{\ln L_z}{L^{d-1}}
	+ x_{\parallel} \frac{\ln L}{L^{d-1}} + \frac{\text{const.}}{L^{d-1}} \quad,
	\end{multline}
$\gamma_{\infty}$ being the interfacial tension in the thermodynamic limit. While the constant in the last term is non-universal (depends on the model and on temperature), the constants $x_{\perp}$ and $x_{\parallel}$ only depend on the ensemble and the boundary conditions~(see table~\ref{tab: ScalingConstants} for numerical values). Translational freedom of the interface(s) in $z$-direction contributes to the first logarithmic term only, capillary waves to the second term, and domain breathing (explained below) contributes to both. In the following, these effects and the values of the universal constants will be motivated and verified by computer simulations in the Ising model in two and three dimensions.

\begin{figure}[htbp]
\centering
\subfigure{\includegraphics[clip=true, trim=5mm 5mm 5mm 5mm, angle=-90,width= 0.99 \columnwidth]{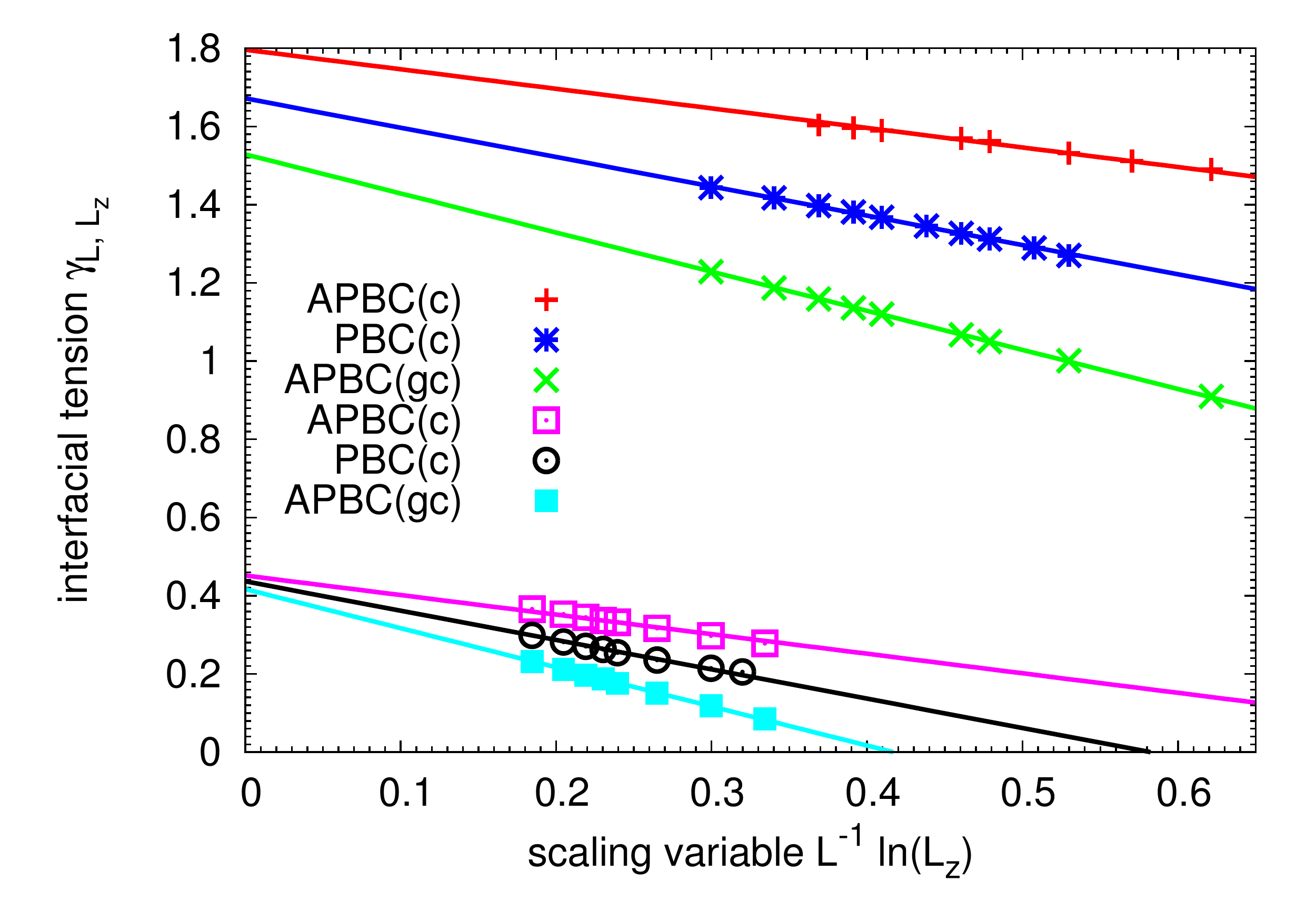}}
\subfigure{\includegraphics[clip=true, trim=5mm 5mm 5mm 5mm, angle=-90,width= 0.99 \columnwidth]{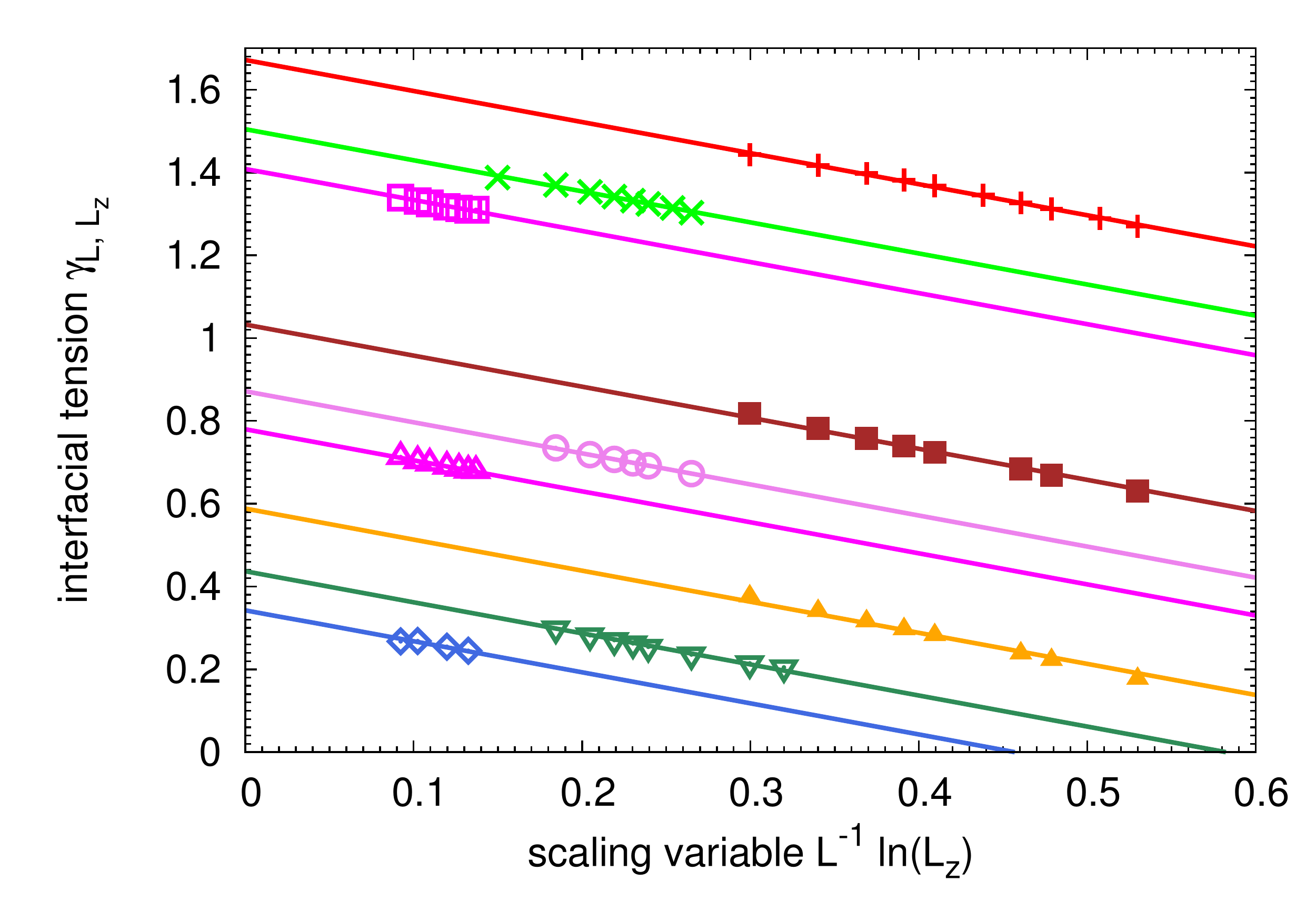}}
\subfigure{\includegraphics[clip=true, trim=5mm 5mm 5mm 5mm, angle=-90,width= 0.99 \columnwidth]{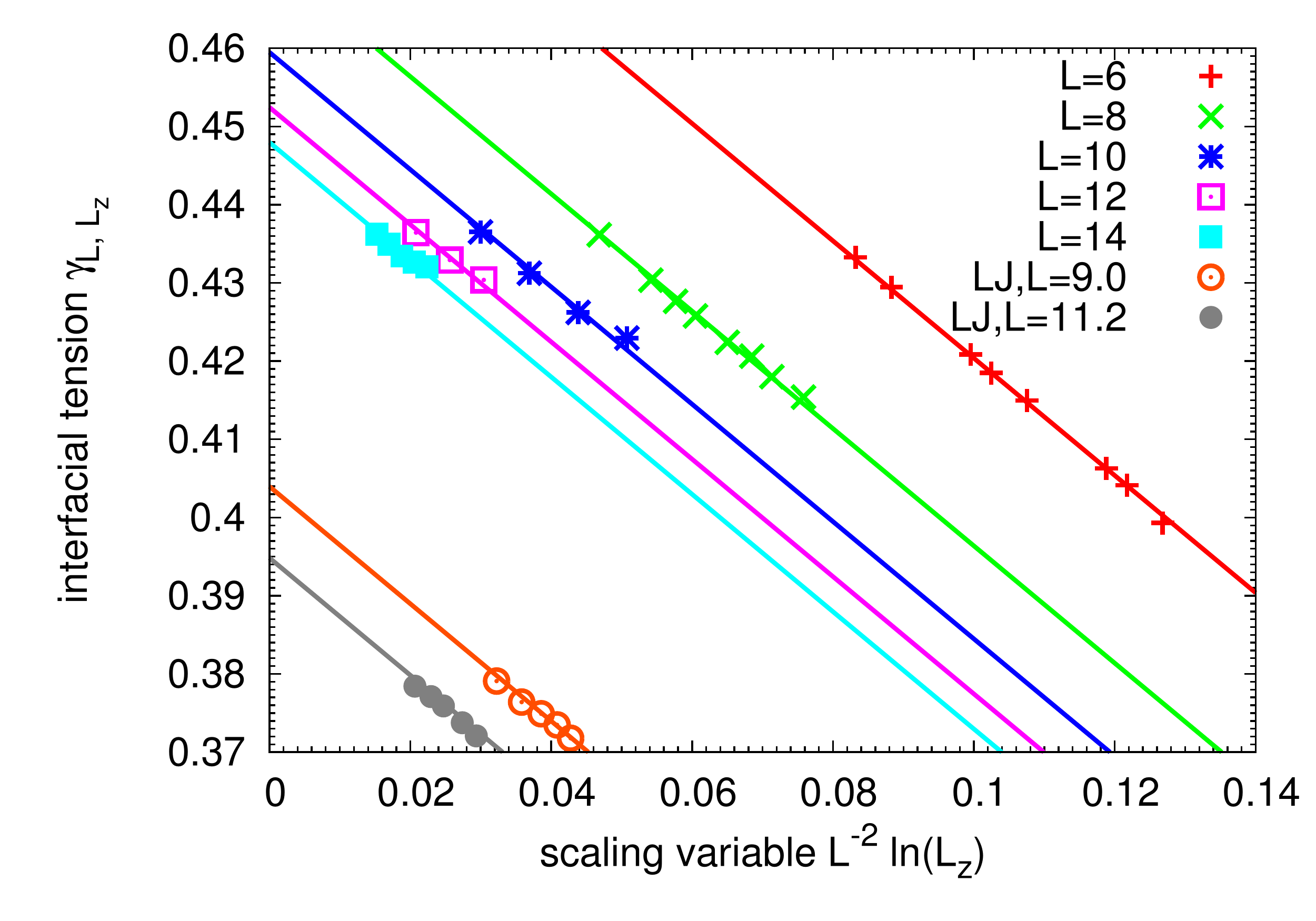}}
\caption{\label{fig3} Interfacial tension $\gamma_{L, L_z}$ for the $d=2$ (a,b) and $d=3$ (c) Ising model, plotted vs. the scaling variable $L^{-(d-1)} \ln L_z$ at fixed $L$. (a) compares the cases APBC(c), PBC(c) and APBC(gc), for two temperatures, $\kb T/J=1.2$, $L=10$, upper three straight lines, and $\kb T/J=2.0$, $L=20$, lower three straight lines. The slopes are theoretical values $x_\perp=1/2$, $3/4$ and $1$, respectively. (b) shows data for PBC(c) in $d=2$ at three temperatures; $\kb T/J=1.2$, three top lines; $\kb T/J=1.6$, three middle lines; $\kb T/J=2.0$, three lower lines; at three choices of $L$ ($L=10, 20, 40$, from top to bottom). These data show that the slope $(x_L=3/4$) neither depends on $L$ nor on temperature. (c) shows data for PBC(c) with $d=3$, $\kb T / J=3$, and 5 choices of $L$, as indicated. (c)~also shows preliminary data for a LJ fluid (see text).}
\end{figure}

The correction with $x_\perp$ is simply interpreted as due to the translational entropy of the interface(s). If an interface is able to move freely in $z$-direction (e.g.~in APBC(gc)), this corresponds to a translational entropy of the form $\kb \ln(L_z)$, yielding $x_\perp=1$ (Fig.~\ref{fig3}). While this is well-known (e.g.~\cite{32,41}), our results for a canonical ensemble with periodic or antiperiodic boundary conditions, where the translational freedom of the interface(s) is constrained, are new. 
We stress that the case APBC(c) is not equivalent to a ``clamped'' interface (with $x_\perp=0$ \cite{41}): When $m=(m_+ + m_-)/2=0$ in Fig.~\ref{fig1}a, still fluctuations of $m_+$, $m_-$ occur, $\delta m_+=m_+-m_\text{coex}, $ $\delta m_- = m_-+m_\text{coex}$. These fluctuations correlate with a fluctuation of the interface position around its mean value. The distance $\Delta$ is found from $m= m_+ (L_z/2-\Delta) + m_- (L_z/2 + \Delta)$ as $\Delta \approx \delta m_+ L_z / (2 m_\text{coex})$. Using for $\delta m_+$ near $m_\text{coex}$ that the probability distribution is a Gaussian $P_{L,L_z/2} (\delta m_+) \propto \exp \{-(\delta m_+)^2 L^{d-1} L_z /(4 \kb T \chi_\text{coex})\}$, $\chi_\text{coex}$ being the susceptibility at equilibrium, we find for this ``domain breathing''
	\begin{equation} \label{eq3}
	\langle \Delta^2 \rangle =\frac{\kb T \chi_\text{coex}}{4 m^2 _\text{coex}} \frac{L_z}{L^{d-1}} \quad.
	\end{equation}
The translational entropy due to fluctuations is $\kb \ln (\sqrt{\langle \Delta ^2 \rangle}/a)$, $a$ being the lattice spacing. In
$\gamma_{L, L_z}$ this yields a correction term $\Delta \gamma$
	\begin{equation} \label{eq4}
	\Delta \gamma = -\frac12 \frac{\ln L_z}{L ^{d-1}} + \left(\frac{d-1}{2} \right) \frac{\ln L}{L^{d-1}} + \frac{\text{const.}}{L^{d-1}} \quad .
	\end{equation}
Thus, for the case APBC(c) the result is $x_\perp=1/2$, as stated in table~\ref{tab: ScalingConstants}. For PBC(c) this result holds with respect to the distance between the interfaces, but the whole positive domain (Fig.~\ref{fig1}b) can freely
translate as a whole. Adding the terms $x_\perp=1/2$ and $x_\perp=1$ for these two degrees of freedom and dividing by the number (2) of interfaces then yields $x_\perp=3/4$. These values are nicely compatible with our numerical results~(Fig.~\ref{fig3}). Preliminary data for a Lennard-Jones (LJ) fluid at $T/T_c=0.78$ ($T_c$ is the vapor-liquid critical temperature) are also included (lengths are in units of the LJ diameter) and compatible with the predicted value of $x_\perp$.

Of course, in \eqref{eq2}, one cannot take the limit $L_z \to \infty$ at fixed $L$. There exists a length $L_{z,0}$ where $\gamma_{L,L_z}$ would become zero: for $L_z>L_{z,0}$ the system can spontaneously break up in multiple domains \cite{11}. Indeed, in the limit $L_z\to \infty$, the typical distance between domain walls is $\xi_\parallel \propto L^{(3-d)/2} \exp(\gamma L^{d-1})$ (the pre-exponential factor is attributed to capillary waves in \cite{37}), and one expects $L_{z,0}$ to be of the same order as $\xi_\parallel$. Our numerical studies (Fig.~\ref{fig3}) have been taken such that $L_z \ll L_{z,0}$.

	\begin{table}[ht]
	\begin{ruledtabular}
	\begin{tabular}{ccccc}
	$d$ & BC & ensemble & $x_\perp$ & $x_\|$ \\
	\hline
	2 & antiperiodic & grandcanonical & $1$ & $1/2$ \\
	3 & antiperiodic & grandcanonical & $1$ & $0$ \\
	2 & antiperiodic & canonical & $1/2$ & $1$ \\
	3 & antiperiodic & canonical & $1/2$ & $1$ \\
	2 & periodic & canonical & $3/4$ & $3/4$ \\
	3 & periodic & canonical & $3/4$ & $1/2$ \\
	\end{tabular}
	\end{ruledtabular}
	\caption{Values of the universal constants in 2 and 3 dimensions. The universal constants only depend on the boundary conditions (periodic or antiperiodic) and the ensemble (canonical or grandcanonical). Note that $x_\perp$ is independent of the dimensionality of the interface while $x_\|$ depends on $d$ because it results from capillary wave effects and \eqref{eq4}.}
	\label{tab: ScalingConstants}
	\end{table}

\begin{figure}[hbtp]
\centering
\subfigure{\includegraphics[clip=true, trim=0mm 0mm 0mm 0mm, angle=-90,width= 0.99 \columnwidth]{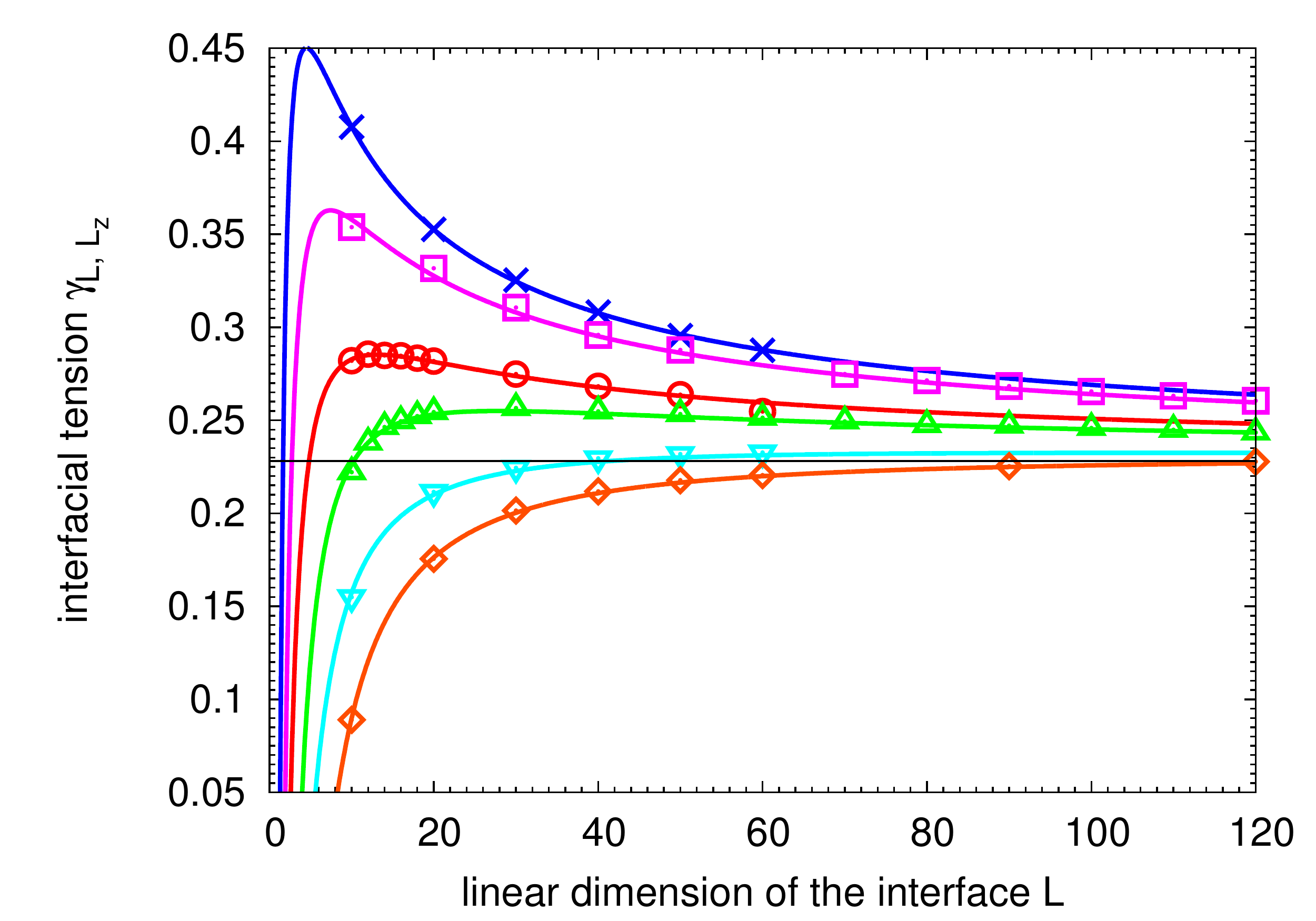}}
\subfigure{\includegraphics[clip=true, trim=0mm 0mm 0mm 0mm, angle=-90,width= 0.99 \columnwidth]{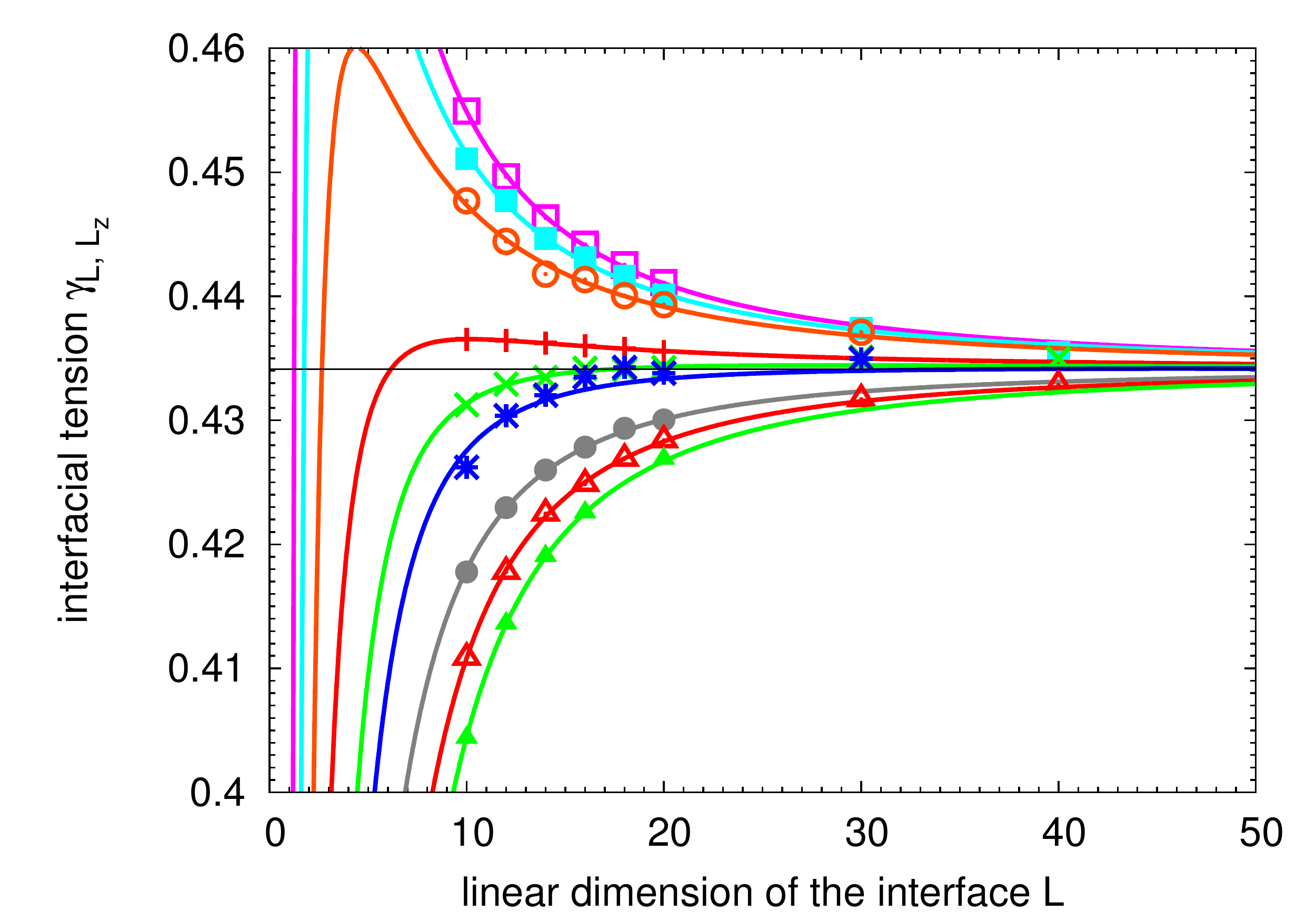}}
\caption{\label{fig4} Interfacial tension $\gamma_{L,L_z}$ plotted vs. $L$. (a) shows data for $d=2, \kb T/J=2.0, L_z=60,120$ while (b) shows $d=3,\kb T/J=3.0, L_z=20,40, 80$. The horizontal straight line shows known values of $\gamma_\infty$ \cite{22,25}, while the curves are fits of Eq.~(\ref{eq2}) to the data (symbols) for the cases APBC(c), top set of curves; PBC(c), middle set; APBC(gc), bottom set; in each set, $L_z$ increases from top to bottom. The theoretical values of $x_\perp$, $x_{\parallel}$ from table~\ref{tab: ScalingConstants} were used.}
\end{figure}

To discuss $x_\parallel$, a correction due to the finite size effect on the capillary waves spectrum has to be taken into account, namely \cite{37} $\frac{3-d}{2} \ln (L) /L^{d-1}$. Ignoring possible $\ln(\ln L)$ corrections right at $d=3$ \cite{49}, one obtains for the APBC(gc) case the values given in table~\ref{tab: ScalingConstants}. For the cases APBC(c) and PBC(c), one has to include $(3-d)/2$ for each interface, but one must also take into account~\eqref{eq4}, resulting in $x_\parallel=(3-d)/2 + (d-1)/2$ for APBC(c) and $x_\parallel=[2(3-d)/2 + (d-1)/2]/2$ for PBC(c), where the overall factor $1/2$ for PBC(c) is due to the two interfaces in the system. The constants for the case PBC(c) also apply for the probability distribution method (Fig.~\ref{fig2}b), different from literature statements, where the above fluctuation mechanism (Eq.~(\ref{eq3})) was missed. Fig.~\ref{fig4} shows excellent agreement with these predictions, both for $d=2$ and $d=3$; note that there is a single constant (from the term $\text{const. } L^{-(d-1)}$ in Eq.~(\ref{eq2})) adjusted in each curve. An important check is that this constant is almost independent of $L_z$, which shows that higher order corrections to Eq.~(\ref{eq2}) are not needed in the cases shown. Also if we let $x_{\parallel}$ as a free parameter, we get results compatible with the theoretical answers, which are summarized in Table~\ref{tab: ScalingConstants}. An interesting aspect is that for large enough $L_z$ the convergence of the APBC(gc) results is from below, while the APBC(c) results converge from above: in the cases of interest, where $\gamma_{\infty}$ is not known in beforehand, this property may give useful bounds on the possible values of $\gamma_\infty$.

An intriguing question is the behavior of systems with continuous spins with antiperiodic boundary conditions~\cite{53}. As is well known, the "interface" then is spread out over the full distance $L_z$, and the free energy cost is not of order $L^{d-1}$ but rather $L^{d-1}/L_z$. While for one-component systems, the interfacial width $w$ depends on $L$ but not on $L_z$, and hence a translational entropy $\ln(L_z/w)$ arises, now $w=L_z$ and hence no term proportional to $\ln L_z$ is expected (and also not found~\cite{53}).

In summary, by discussing the interfacial tension in finite systems as a function of both linear dimensions $L$ and $L_z$ (unlike large parts of the previous simulation literature which focused on $L=L_z$) we have identified the mechanisms of the finite size corrections. The knowledge of these corrections allows to obtain more reliable estimates of the interfacial tension in the thermodynamical limit. A crucial point is the comparison of different boundary conditions (periodic or antiperiodic) and ensembles (canonical or grandcanonical). While the numerical examples are mostly from the Ising model, we stress that a fixed spin boundary condition at $z=0$ and $z=L_z$ gives (in the Ising model) results fully equivalent to the APBC case, both for the grandcanonical and canonical ensembles. This can be easily generalized to arbitrary systems; e.g.~for a study of solid-liquid interfaces one needs to choose boundary potentials that stabilize the solid on one wall and the liquid on the other wall. Of course, in such cases it is already a nontrivial matter to identify precisely the conditions where phase coexistence occurs in the bulk. Nevertheless, we expect that our analysis will be useful for studies of many model systems, and will also help to understand possible experiments on interfacial phenomena in nano-confinement. We also mention that our treatment can be extended to understand finite size effects on droplet free energies, hampering the estimation of Tolman's length \cite{51,52} that describes curvature corrections to the surface free energy of droplets.

\begin{acknowledgments}
This research was supported by the Deutsche Forschungsgemeinschaft (DFG), grant No VI $237/4-3$. One of us (K.B.) acknowledges stimulating discussions with A. Tr\"oster and M. Oettel.
\end{acknowledgments}


\begin{thebibliography}{99}
\bibitem{49} Capillary wave theory yields the interfacial width as $w\propto L^{(3-d)/2}$ in $d<3$ but $w\propto \sqrt{\ln L}$ in $d=3$. Hence it is possible that the correction $L^{-(d-1)} \ln[ L^{(3-d)/2}]$ should be interpreted as $L^{-2} \ln [ \sqrt{\ln L} ]$ in $d=3$.
\bibitem{52} For a droplet with surface area $A$ in a cubic box of volume $L ^3$ we would predict a correction $-(3 \ln L)/A$ from its translational entropy and $-(3/2) (\ln L) / A + (\ln A)/A$ from the ``breathing mode'', analogous to Eq.~(\ref{eq3}).
\end{thebibliography}
\end{document}